\documentclass[preprint,authoryear,12pt]{elsarticle}

\usepackage{amssymb}

\usepackage[polutonikogreek,english]{babel}
\usepackage[latin1]{inputenc}

\journal{Journal of Computational Science}

\begin{document}

\begin{frontmatter}



\title{Mixed Reality Serious Games: The Therapist Perspective}


\author{Di Loreto Ines, Goua\"{\i}ch Abdelkader}

\address{LIRMM, Universit\'e Montpellier 2-CNRS, France}

\begin{abstract}
The objective of this paper is to present a Mixed Reality System (MRS) for rehabilitation of the upper limb after stroke. The system answers the following challenges:
(i) increase motivation of patients by making the training a personalized experience; (ii) take into account patients' impairments by offering intuitive and easy to use interaction modalities; (iii) make it possible to therapists to track patient's activities and to evaluate/track their progress;  (iv) open opportunities for telemedicine and tele rehabilitation; (v) and provide an economically acceptable system by reducing both equipment and management costs.
In order to test this system a pilot study has been conducted in conjunction with a French hospital in order to understand the potential and benefits of mixed reality. The pilot involved 3 therapists who 'played the role' of  patients. Three sessions, one using conventional rehabilitation, another using an ad hoc developed game on a PC, and another using a mixed reality version of the same game were held.  Results have shown the  MRS and the PC game to be accepted more than physical rehabilitation.
\end{abstract}


\end{frontmatter}



\section{Introduction}
\label{Introduction}
In just the United States each year approximately 795.000 people are affected by stroke \citep{diseaseUSA}. In France strokes are around 150.000 each year \citep{diseaseFR}. Similar trends are shown all around the industrialized world. A stroke can cause disabilities, such as paralysis, speech difficulties, and emotional problems. These disabilities can significantly affect the quality of daily life for survivors. Several researchers on the field have shown that important variables in relearning motor skills and in changing the underlying neural architecture after a stroke are the quantity, duration, and intensity of training sessions \citep[for more information see e.g.,][]{Micera2005}. 

 In this paper we  present a mixed reality approach for upper limb rehabilitation focused  to increase quantity, duration, and quality of training sessions. In particular the approach answers the following challenges:
(i) increase motivation of patients by making the training a personalized experience, for example adjusting exercises difficulty according to patients' cognitive and physical capabilities; (ii) take into account patients impairments by offering intuitive and easy to use interaction modalities that do not require a learning curve and that are adapted to elder people; (iii) make it possible to therapists to track patient's activities and evaluate progress in order to  adapt therapeutic strategies; (iv) open opportunities for telemedicine and tele rehabilitation when patients leave the hospital; (v) provide an economically acceptable system by reducing both equipment and management costs.

It is worth to note that the above mentioned issues address two targets: the therapist and the patient. It is our opinion that  it is important to evaluate acceptance of these systems from both, the patient's point of view and the therapists' point of view. Otherwise, even if clinical efficiency is demonstrated, developed systems will be used only for academic studies and may not be widely accepted in real rehabilitation centers. 
For this reason, before testing the system directly with patients we decide to conduct a pilot study in conjunction with a French hospital. Aim of this pilot was understand the potential and benefits of mixed reality from the therapist's point of view. The pilot involved 3 therapists who 'played the role' of  patients. Three sessions, one using conventional rehabilitation, another using an ad hoc developed game on a PC, and another using a mixed reality version of the same game were held. 
Before entering in detail into the pilot study description, in following subsections we will describe in more depth what a stroke is and what are its consequences, and why a 'virtual'  approach could be useful for such a rehabilitation.

\subsection{Describing a stroke}
\label{stroke}
After a stroke there is a loss of brain functions due to disturbance in the blood supply to the brain. The affected area of the brain is unable to function, leading to inability to move one or more limbs on one side of the body and perhaps to cognitive problems such aphasia or the so-called neglect effect. Because of these impairments stroke sufferers are often unable to independently perform day-to-day activities such as bathing, dressing, and eating. Nearly three-quarters of all strokes occur in people over the age of 65. The specific abilities that will be lost or affected by stroke depend on the extent of the brain damage and most importantly where in the brain the stroke occurred. For example, a stroke in the right hemisphere often causes paralysis in the left side of the body. This is known as left hemiplegia. Survivors of right-hemisphere strokes may have problems with their spatial and perceptual abilities. This may cause them to misjudge distances (leading to a fall) or be unable to guide their hands to pick up an object, button a shirt or tie their shoes. They may even be unable to tell right-side up from upside-down when trying to read. Survivors of right-hemisphere strokes may also experience left-sided neglect. Stemming from visual field impairments, left-sided neglect causes the survivor of a right-hemisphere stroke to 'forget' or 'ignore' objects or people on their left side. 
On the other hand, someone who has had a left-hemisphere stroke may develop aphasia. Aphasia is a catch-all term used to describe a wide range of speech and language problems. These problems can be highly specific, affecting only one component of the patient's ability to communicate, such as the ability to move their speech-related muscles to talk properly. The same patient may be completely unimpaired when it comes to writing, reading or understanding speech. By contrast to survivors of right-hemisphere stroke, patients who have had a left-hemisphere stroke often develop a slow and cautious behavioral style. They may need frequent instruction and feedback to complete tasks  \citep[information from the][]{strokeEffects}.

As we can see from these two examples, the consequences of  two stroke accidents could be deeply different. For this reason each stroke rehabilitation program is personal, designed for a particular patient, and not a generic one. 

\subsection{Strokes and rehabilitation}
\label{theraphy}
In stroke accidents rehabilitation involves intensive and continuous training to regain as much function as possible, depending on several factors including the severity of brain lesions and the degree of cerebral plasticity. 
Following the hypothesis that the theories of motor learning can be applied on motor relearning most rehabilitation techniques are founded on principles of motor learning and skill acquisition established for the healthy nervous system.
Acknowledged features are among others: (i) the motivation of the participant; (ii) the use of variable practice (i.e. practice a variety of related tasks); (iii) training with high intensity/many repetitions; (iv) and providing feedback  \citep{Krakauer06, Levin, Langhorne}. 

In addition, therapy on the lower extremity (i.e., legs) is the primary concern in early inpatient stroke therapy in order to enable mobility of the patient. Recovery of the upper extremity (i.e., arms) has a slower progression and is usually gained through outpatient and home therapy \citep{Lee99}. Patients with upper extremity paralysis typically regain motion starting from their shoulder. Over time, they may gradually regain motion in the elbow, wrist, and, finally, the hand. The most important part of stroke rehabilitation is conducted during the first 6 months after the stroke and due to the cost, only 6/8 weeks of rehabilitation is done under the continued direct supervision of an expert (i.e., in the hospital). Because of limitations on therapy patients must do much of the work necessary to recover arm function at home.

To summarize, after a stroke the affected area of the brain is unable to function, leading to inability to move one or more limbs on one side of the body and also perhaps to cognitive problems such aphasia or the so-called neglect effect. As a result of these impairments stroke sufferers are often unable to independently perform day-to-day activities such as bathing, dressing, and eating. As a side effect they can develop depression or aggressiveness due to the trauma of reduced capabilities. Depression and aggressiveness also imply that stroke survivors may find it difficult to focus on a therapy programme.  In addition, while these programmes attempt to stimulate the patient with a variety of rehabilitation exercises, stroke victims commonly report that traditional rehabilitation tasks can be boring due to their repetitive nature. We can then say that motivation is an important factor for rehabilitation success.  

\subsection{Increasing rehab volume}
\label{motivation}

Several researchers have shown, both in animal and human, that important variables in relearning motor skills and in changing the underlying neural architecture are the quantity, duration, and intensity of training sessions.
In particular, research in animal models suggests that with intensive therapy (repeating individual motions hundreds of times per day), animals that experience strokes can recover a significant amount of their lost motor control \citep{Selzer}.  Similarly, recent guidelines for treatment of human patients recommend high-intensity, repetitive motions while keeping patients informed about their progress \citep{Langhorne}. Looking at the effects of different intensities of physical therapy treatment, several authors \citep{Langhorne96, Taub99,Taub00} have reported significant improvement in activities of daily living as a result of higher intensities of treatment. 

To experience significant recovery, stroke patients must perform a substantial number of daily exercises. Unfortunately, typical sessions with therapists include a relatively small number of motions \citep{Lang}. One of the possible solutions is doing rehabilitation at home. However, while therapists prescribe a home exercise regimen for most patients, a study indicates only 31\% of patients actually perform these exercises as recommended \citep{Shaughnessy}. Thus, home based stroke rehabilitation has the potential to help patients in recovering from a stroke. On the other hand a  main problem arises when choosing  which kind of home-based technology can both help and motivate patients to perform therapeutic exercises.

\section{Motivation: Advantages of 'virtual rehabilitation'}
\label{advantages}
The challenge for post stroke rehabilitation is to create exercises able to decrease monotony of hundreds of repeated motions. In order to overcome the above mentioned problems different kinds of 'non traditional' therapies have been proposed. 
For example, the possibility of using 'virtual' rehabilitation has been the subject of experiments by several authors (for example \citep{Rizzo2005,burdea}). Although most of studies on this topic are linked to the study of virtual reality environments recent studies have focused on the use of videogames and consoles for rehabilitation (such as \citep{Burke2009c,Flynn}. 

Results of these studies can be summerized as follows: 
\begin{enumerate}
  \item	\emph{Personalization}: Virtual rehabilitation technology creates an environment in which the intensity of feedback and training can be systematically manipulated and enhanced in order to create the most appropriate, individualized motor learning paradigm  \citep{burdea}. Rehabilitation using games can take advantage of the use of adaptation in order to create ad hoc personalized games.
  \item \emph{Interactivity}: An advantage present in all forms of virtual rehabilitation is the use of interactivity. For example it has been suggested that integrating gaming features in virtual environments for rehabilitation could enhance user motivation. Virtual rehabilitation exercises can be made to be engaging, such that the patient feels immersed in the simulated world. This is extremely important in terms of the patient motivation \citep{Popescu}, which, in turn, is key to recovery.  A person who enjoys what he is doing spends more time developing her skills in a given activity.
  \item	\emph{Feedback}: Interactive feedback can contribute to motivation. For example, by providing visual and auditory rewards, such as displaying gratifying messages in real time, patients are motivated to exercise \citep{Chen2006b,Goude2007a}
  \item	\emph{Tracking}: The evolution of the patient's performance can be easily stored, accessed and developed without the patient's or therapist's input. In addition, the Internet can be used for data transfer, allowing a therapist to remotely monitor progress and to modify the patient's therapy program \citep{burdea, Popescu}. 
  \item \emph{Telerehabilitation}: Virtual rehabilitation can stimulate the patient using a variety of rehabilitation exercises at a low cost. This means that rehabilitation costs can be contained if the technology used for virtual rehabilitation (consoles and games) is easily accessible.  In addition lower cost personal equipment (for example pc-based) will eventually allow rehabilitation stations to be placed in locations other than the rehabilitation center, such as a patient's home.
 \end{enumerate}

On the other hand, virtual rehabilitation raises important challenges that may limit its widespread adoption such as:
\begin{itemize}
  \item	Clinical acceptance, which relies on proved medical efficacy
  \item	Therapist's attitude towards the technology (e.g., the therapist agrees that  technology  is able to replace therapists and the like)
  \item	Patient's attitude towards the technology (e.g., the patient may not consider a game to be 'real' rehabilitation).
  \item	Challenges linked to the kind of technology used (challenges differ from virtual reality to consoles). 
 \end{itemize}

\subsection{Acceptance}
\label{acceptance}

\begin{figure}[h]
	\centering
		\includegraphics[scale=0.5]{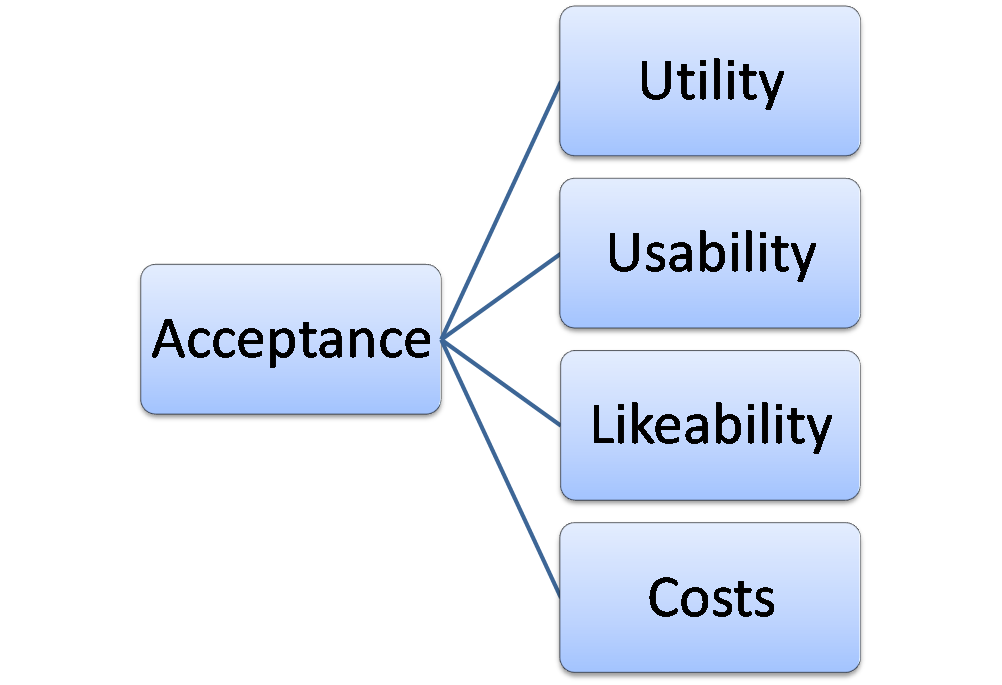}
	\caption{Shackel's definition of usability: first level}
	\label{fig:shackel}
\end{figure}

In order to overcome the last three challenges we propose an approach based on Mixed Reality (see section \ref{sytems} of this paper for the description of such a system).  However, while evaluating usability for systems has become a standard \citep{Nielsen, Macleod94, Bevan95} attitude is a more complex element which requires an in depth assessment. 
For this reason we propose to use an approach meant to evaluate 'acceptance' of a tool based on Shackel.
Shackel \citep{Shackel1991} defines a model where product acceptance is the highest concept. The user has to make a trade off between utility, the match between user's needs and functionality, usability, ability to use functionality in practice and likeability, affective evaluation vs costs (see Fig.\ref{fig:shackel})

It is worth noting that in a system for stroke rehabilitation we are dealing with two kind of acceptances. 
The first one is patient's acceptance of the system. In order to be accepted by our users  the system has to be considered usable, useful and likeable. The second one  is  therapist's acceptance.  In addition to the above mentioned elements the system has also to show adequate cost for the therapist. While in this paper we address directly the therapist acceptance and the patient's perspective only as a simulate one (see Section \ref{protocol}) it's our opinion that both perspectives are very important.

\section{State of the art}
\label{relworks}
This section reviews existing approaches for post stroke rehabilitation. The presentation of works is not exhaustive and the focus has been made on three types of rehabilitation systems: robot based system, virtual reality based systems and mixed reality based systems. Besides, only early stages of rehabilitation have been considered which excluded for instance system dedicated to the rehabilitation of fingers such as \citep{Cameirao2010}.

To avoid ambiguities, hereafter definition of each type of system is given.
\paragraph{Definition of virtual reality}

Virtual reality system is commonly defined as an environment that can simulate real world situations and allows users to interact within this simulated environment using different types of devices such as head mounted device (HMD) and gloves.

\paragraph{Definition of mixed reality}

Mixed reality system merges real and virtual worlds to create a new context of interaction where both physical and digital objects co-exist and interact consistently in real time. This contrasts with virtual reality where the interaction is directional from the physical world where the user is situated towards the virtual environment where the interaction holds. Within mixed reality system, the interaction holds among objects of both virtual and physical environments. A classical example of a mixed reality game is the MR pong \citep{Kiia2001}  where players play Pong game with a virtual ball colliding with physical objects on a table.

\paragraph{Definition of robot assisted rehabilitation}

Robot based rehabilitation uses a training robot. The robot complements or induces patient's movement and provides feedback. Robot systems are often coupled with a computer program used to create virtual context for actions and delivering a visual feedback.

\subsection{Evaluation framework}

An evaluation framework is used to analyze usability of each system from therapist, patient and financier point of view.

\subsubsection{Therapist perspective}

From the therapist's point of view, we are interested by the following evaluation criteria:
\par
\paragraph{Therapist intervention}
\par
This criterion informs whether the therapist can intervene during the rehabilitation sessions. In fact, we have noticed from observations and preliminary interviews with therapists the importance of assistance and guidance delivered by the therapist. The therapist has to intervene physically during the rehabilitation sessions to support the patient and to prevent him from developing incorrect compensating gestures such as chest balancing. Possible values of this criterion are: yes, if the therapist can intervene and no, otherwise.
\par
\paragraph{Changes on therapist habits}
\par
This criterion informs about changes induced by introduction of the rehabilitation system to therapist working habits. The introduction of a new rehabilitation system may induce changes in the way therapists were conducting therapeutic sessions. The goal of this criterion is to evaluate the amount of the induced changes. Possible values are: negligible, if the therapists has not to change, her usual way of work; moderate, when the therapist has to change some of her working habits and important, when the therapist has to change her way of conducting therapies.  
\par
\paragraph{System setup}
\par
This criterion informs about the setup phase of the system. Often, the rehabilitation system needs some setup phase in order to get devices installed and configured correctly before starting rehabilitation sessions. This criterion informs whether the therapist alone can perform this setup phase or a specialized assistant is required to setup the system. Possible values of this criterion are: therapist, if the therapist could perform the setup phase and assistant, if intervention of a specialized technician is required.
\par
\paragraph{Location}
\par
This criterion concerns the place where the rehabilitation sessions can be performed. In fact, for some systems such as robot based systems and some virtual reality simulators a special room has to be dedicated to perform rehabilitation sessions. Other lightweight systems do not require a fixed infrastructure and can be used almost anywhere.

\subsubsection{Patient perspective}
From the patient's point of view, we are interested in two evaluation criteria that are:
\par
\paragraph{Eye-Hand focus}
\par
Observations from therapeutic sessions and interviews with therapists have revealed an important point related to the interaction of the patient with the system. In fact, post stroke patients during early stages may be reluctant to any system that requires an important cognitive effort. Ideally, the attention of the patient has to be attracted and focused on a single point. For instance, when given a mouse patients look at their hand when trying to move and not to the screen. Consequently, they find it difficult to follow two actions at the same time: the hand that is moving and the game displayed on the screen.
This criterion evaluates whether the eye and hand of the patient are attracted to the same place or to different places. Possible values for this criterion are: same place, when the hand and eyes of the patient are directed to the same place and different places, when the hand and eyes of the patient are attracted to different spaces.
\par
\paragraph{Invasiveness of the system}
\par
This criterion informs about the invasiveness of the system from the patient's perspective. In fact, systems may require the patient to wear special devices to track their movement or to deliverer feedback. These devices are more or less convenient to wear and to tolerate by patients especially the ones that suffer from impairment and muscular spasticity. Possible values for this criterion are: convenient, if the system is considered as convenient to use; invasive, if the system is considered as interfering with the patient.

\subsubsection{Economical perspective}
For a rehabilitation system to be used in rehabilitation centers, it is necessary to demonstrate its medical efficiency and also to show that the system is economically sustainable. The economical aspect of the rehabilitation system are evaluated through two criteria that are:
\par
\paragraph{Unitary cost of the system}
\par
This criterion evaluates the unitary cost of the system in kilo euros (KE). The objective is not to deliver a precise price, but evaluate an order of magnitude of each system cost. Thus, possible values of this criterion are: less than 1 KE per unit, 1-5 KE per unit, 5-10 KE per unit, and more than 10  KE per unit.
\par
\paragraph{Extra required resources}
\par
While the previous criterion estimates the unitary cost of the system, this criterion evaluates all extra resources that must provided for running the system such as recruiting specialized personnel, making available a special and so on.

\subsection{State of the art review}  
\par
\subsubsection{Robot assisted rehabilitation with Manus}

\begin{figure}[h]
	\centering
		\includegraphics[scale=0.7]{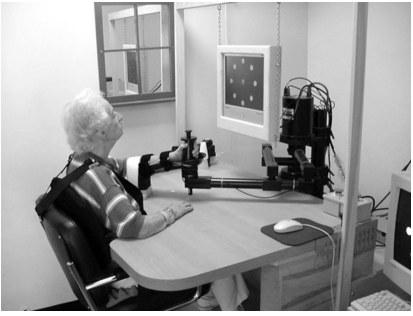}
	\caption{Patient using the Manus Robot \citep{Volpe2001}}
	\label{fig:manus}
\end{figure}

Robot based approach for rehabilitation has been used since the late 90s to assist patients. The Manus Robot developed at the MIT \citep{Volpe2001} is an exemplar system (Figure \ref{fig:manus}). In this system the patient is in front of the robot and her shoulder is strapped to a chair. The patient's impaired arm is strapped to a wrist carrier and attached to the manipulandum. A video screen is above the training table to create a virtual context for movements and provide visual feedbacks to the patients. Several clinical trials have been conducted to evaluate the efficiency of robots based rehabilitation when compared to classical therapies. For instance, in \citep{LoAlbertC2010} it has been demonstrated that robot-assisted therapy improved outcomes for long period of training when compared with usual care.

\begin{table}[h]
\caption{Therapist perspective}
\center
\label{tab:theraper}
\begin{tabular}{|l|l|l|l|l|}
\hline
& Intervention & Habit change & System setup &	Location \\
\hline
\citep{LoAlbertC2010} & Important &  Yes	 & Assistant	 & Dedicated\\
\hline
\end{tabular}
\end{table}

Since the patient is attached to the robot, the therapist cannot intervene directly to assist or provide guidance. All its interventions are mediated by the robot, which induces a big change in the way usual therapies are conducted. The setup and maintenance of the robot requires an advanced knowledge that could be performed by a specialized operator. Due to the unitary cost and setup complexity, therapies with the robot are intended to be performed in dedicated places.

\begin{table}[h]
\caption{Patient Perspective}
\center
\label{tab:pataper}
\begin{tabular}{|l|l|l|}
\hline
& Eye-Hand  Focus & Invasiveness \\
\hline
\citep{LoAlbertC2010}  & Different Places &  Yes\\
\hline
\end{tabular}
\end{table}

With the Manus robot the patient moves her arm and perceives actions on a video screen put in front of her. This makes the eyes and hand acting on different places. The robot is an invasive system requiring that the patient arm and chest to be strapped.

\begin{table}[h]
\caption{Financier Perspective}
\center
\label{tab:finaper}
\begin{tabular}{|l|l|l|}
\hline
& Unitary Cost & Extra resources \\
\hline
\citep{LoAlbertC2010}  &  More 10KE &  Yes\\
\hline
\end{tabular}
\end{table}

The unitary cost of the Manus robot has an order of magnitude about 10 KE and may require extra resources such as recruiting technicians to facilitate management and setup of the robot.
 
\par
\subsubsection{Virtual reality with IREX}
\par
\begin{figure}[h]
	\centering
		\includegraphics[scale=0.3]{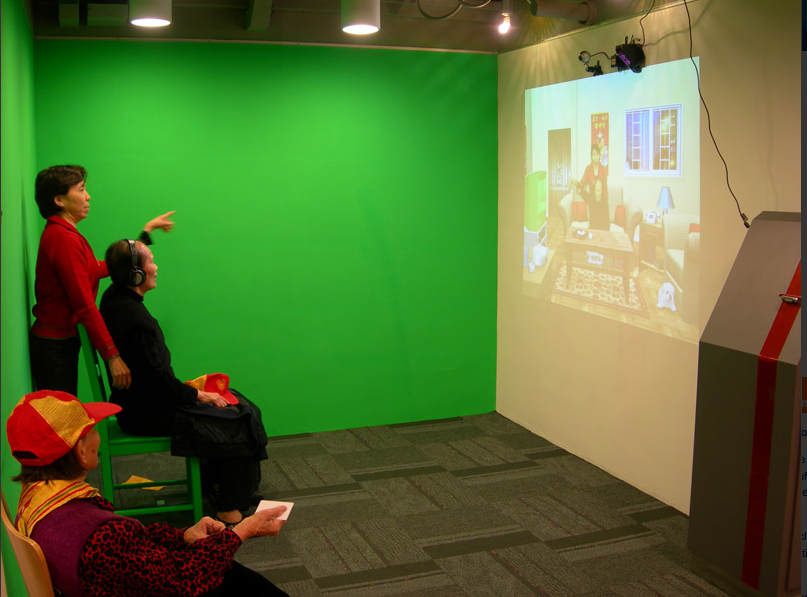}
	\caption{IREX system a projected image on a monitor (www.gesturetekhealth.com) }
	\label{fig:manus}
\end{figure}

The IREX VR \citep{IREX} is a projected video-capture device that has been used and evaluated in several rehabilitation systems \citep{Kizony2005,Reid2004}.
Within this system participants are placed in front or a chroma key green scene and their image is merged, at realtime, with a virtual environment and projected on a monitor. A camera is used for tracking participant's movements and converting them to interaction events within the virtual environment. Games on this system include soccer games, ski and touching balls.

\begin{table}[h]
\caption{Therapist perspective}
\center
\label{tab:theraper}
\begin{tabular}{|l|l|l|l|l|}
\hline
& Intervention & Habit change & System setup &	Location \\
\hline
\citep{Kizony2005} & & & &\\
\citep{Reid2004} & Yes &  Moderate	 & Technician	 & Dedicated\\
\hline
\end{tabular}
\end{table}

As reported by \citep{Kizony2005} the therapist can intervene during the therapeutic sessions. When presenting this system to therapists, some of them have reported the fact that their way of work could be modified since they have to control both the patient and the monitor at the same time to see what is happening in the virtual world.

The setup of the IREX system requires IT competences of a technician to adjust video and motion capture systems. Since, the IREX requires a special chroma key green scene and a bright ambient light, rehabilitation sessions must be performed only in dedicated places.
 
\begin{table}[h]
\caption{Patient Perspective}
\center
\label{tab:pataper}
\begin{tabular}{|l|l|l|}
\hline
 & Eye-Hand  Focus & Invasiveness \\
\hline
\citep{Kizony2005} & &\\
\citep{Reid2004} & Different Places &  No\\
\hline
\end{tabular}
\end{table}

The patient sees her image on a monitor, which implies an additional cognitive effort to situate herself in the virtual world. No invasiveness is reported for IREX since the patient is free from any device.

\begin{table}[h]
\caption{Financier Perspective}
\center
\label{tab:finaper}
\begin{tabular}{|l|l|l|}
\hline
& Unitary Cost & Extra resources \\
\hline
\citep{Kizony2005} & &\\
\citep{Reid2004}& 5-10KE	 &  Yes\\
\hline
\end{tabular}
\end{table}

Finally, the IREX system cost is estimated between 5-10 KE and may require additional resource such as an operator to setup and manage the IT system.

\par
\subsubsection{Virtual reality with Sony EyeToy}

The Sony EyeToy has been used as an economic alternative to IREX for VR rehabilitation \citep{Rand2004}. EyeToy is an off-the-shelf low cost gaming device that allows interaction with virtual objects displayed on a standard TV. By contrast to IREX, EyeToy requires neither a chroma key scene nor a special ambient light.

\begin{table}[h]
\caption{Therapist perspective}
\center
\label{tab:theraper}
\begin{tabular}{|l|l|l|l|l|}
\hline
& Intervention & Habit change & System setup &	Location \\
\hline
\citep{Rand2004} & Yes &  Moderate	 & Therapist	 & Anywhere\\
\hline
\end{tabular}
\end{table}

Similarly to IREX, the therapist can intervene during sessions to assist and provide guidance to patients. However, the therapist has to control both the patient and the monitor at the same time. By contrast to IREX, EyeToy does not require chroma key scene so it can be deployed anywhere opening opportunities for telerehabilitation.

\begin{table}[h]
\caption{Patient Perspective}
\center
\label{tab:pataper}
\begin{tabular}{|l|l|}
\hline
 Eye-Hand  Focus & Invasiveness \\
\hline
Different Places &  No\\
\hline
\end{tabular}
\end{table}

The patient is free from any device which makes EyeToy non invasive. However, similarly to IREX, the patient has to make an additional cognitive effort to situate herself in the virtual environment to coordination her arms and interaction with virtual objects.

\begin{table}[h]
\caption{Financier Perspective}
\center
\label{tab:finaper}
\begin{tabular}{|l|l|}
\hline
 Cost & Extra resources \\
\hline
$<$ 1 KEs &  No\\
\hline
\end{tabular}
\end{table}

EyeToy is a low-cost device available off-the-shelf. A set of commercial games have been used and evaluated experimentally and proved to improve results of post-stroke rehabilitation \citep{Yavuzer2008}. EyeToy system was originally made for entertainment and used with Sony PlayStation. Consequently, using this system does not require special competences.

\par
\subsubsection{Mixed Reality with Tabletop}
\par

\begin{figure}[h]
	\centering
\includegraphics[scale=0.7]{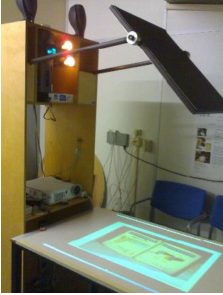}
\caption{Tabletop VIP system \citep{AlMahmud2008}}
	\label{fig:vip}
\end{figure}
\citep{AlMahmud2008} have developed a social game using a tabletop augmented reality platform, namely Visual Interaction Platform (VIP) \citep{Aliakseyeu2001}. VIP allows interaction with physical objects that are tracked using infrared based system. A visual feedback is provided by projecting the display on the table (Figure \ref{fig:vip}) It is worth noting that this system has been initially used for socializing elder people. Nevertheless, the VIP could be adapted for post stroke rehabilitation.

\begin{table}[h]
\caption{Therapist perspective}
\center
\label{tab:theraper}
\begin{tabular}{|l|l|l|l|l|}
\hline
& Intervention & Habit change & System setup &	Location \\
\hline
\citep{AlMahmud2008} & Yes &  Negligeable	 & Assistant	 & Dedicated\\
\hline
\end{tabular}
\end{table}

Since the patient is free from any device the therapist can intervene during sessions and consequently the system does not implies changing usual ways of work. The VIP device is a home made prototype so it requires presence of an operator to setup the system. Besides, VIP is a quite large system so a dedicated place is required.

\begin{table}[h]
\caption{Patient Perspective}
\center
\label{tab:pataper}
\begin{tabular}{|l|l|l|}
\hline
& Eye-Hand  Focus & Invasiveness \\
\hline
\citep{AlMahmud2008} & Same Place &  No\\
\hline
\end{tabular}
\end{table}

The patient interacts with her hand with objects on the table were the display is projected. This makes interaction very natural with both eyes and hand are targeting the same place. Since no device is attached to the patient, the VIP is considered as not invasive.

\begin{table}[h]
\caption{Financier Perspective}
\center
\label{tab:finaper}
\begin{tabular}{|l|l|l|}
\hline
& Unitary Cost & Extra resources \\
\hline
\citep{AlMahmud2008} & n-a &  n-a\\
\hline
\end{tabular}
\end{table}

VIP device used in \citep{AlMahmud2008} is a home made prototype, so no information is available to estimate its cost.

\subsection{Discussion:}  

Different devices have been used for post stroke rehabilitation ranging from robots, video projection and mixed reality systems. These systems have different characteristics when regarded from therapist, patient and financier perspectives.

Robot assisted rehabilitation shows interesting results as demonstrated by large clinical trials \citep{LoAlbertC2010}. These results are explained by the amount of movements produced by the robots and executed by the patient.
Using a robot prevents the therapist from direct intervention with patient which may causes important changes on usual practices of rehabilitation and may reduce consequently acceptance by therapists. Robot are also invasive devices and the eye-hand coordination is not facilitated in systems like Manus since the action's feedback is provided on a monitor placed in front of the patient and not directly on the physical world.  

By contrast to robot-based systems, virtual reality rehabilitation using IREX or EyeToy devices allows intervention of the therapist. By contrast to IREX, that is a costly system requiring a special setup with a chroma key scene, EyeToy is a low cost device usable anywhere. However, games used by EyeToy are not targeting post stroke patients and consequently don't take into account patient's impairments caused by the stroke. Patient's eyes and hands are attracted to different regions in both systems.

By contrast to IREX and EyeToy based systems, within mixed reality systems such as the one using VIP both the eyes and the hands of the patient are attracted to the same region, which reduces the cognitive effort and makes interaction more natural. However, the VIP platform is a homemade prototype that could not be used in real rehabilitation centers.

It is worth noting, that several studies have analyzed the efficiency from patient's perspective by measuring recovery metrics such as Fugl-Meyer index \citep{P.W.DuncanM.Propst1983}.  In our opinion, parallel to these studies it is important to evaluate the acceptance of these systems from the therapists' point of view. Otherwise, even if clinical efficiency is demonstrated, developed systems will be used only for academic studies and may not be widely accepted in real rehabilitation centers. This has motivated our study by presenting the point of view of therapist on a mixed reality system by asking them to play the role of patients.


\section{The Mixed Reality System}
\label{methods}
In order to  test the Mixed Reality System we created, a pilot study was carried out with therapists, asking them to 'play the role' of the patients. This choice was driven by the idea that - because of their years spent with patients - they could be able to 'simulate' patients' reactions to the system. 
In particular aim of this experiment was to compare the use of an in-house designed Mixed Reality System (MRS) with two alternative single-user
tools through an empirical investigation. The two alternative tools are classical physical rehabilitation, and a
an ad-hoc post stroke PC game. We expected the  MRS and the PC game to be accepted more than physical rehabilitation by the patient, and MRS easiness of use to be considered higher than the PC one.  The pilot study was not intended to deliver statistical evidence, but simply to give approximate values guiding the set-up of following experiments. 

\begin{table}[h]
\caption{Hypothesis for the pilot study}
\center
\label{tab:ipotesi}
\begin{tabular}{|l|l|}
\hline
H1: & MRS and the PC game acceptance is higher than \\& physical rehabilitation acceptance \\
\hline
H2: & MRS easiness of use is evaluated higher than the PC one \\
\hline
\end{tabular}
\end{table}

Hereafter the MRS is described in detail. 

\subsection{The system}
\label{sytems}
As said previously, a mixed reality game aims at merging real and virtual worlds to create a new context of interaction where both physical and digital objects co-exist and interaction consistently in real time. Our objective in this study is to build such a system for rehabilitation games taking into account the following characteristics:
\begin{itemize}
\item easiness of use and management: both hardware and software components of the system must be easy to setup and install without inducing important management and maintenance costs.
\item reduced cost: the system has to be made of commercially available components that are affordable to any institution aiming to reproduce and use the therapeutic games.
\end{itemize}
\par\medskip
\textbf{Description of the system}

\begin{figure}[h]
	\centering
		\includegraphics[scale=0.6]{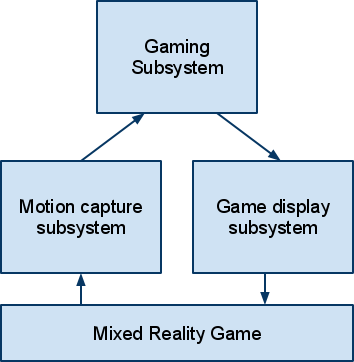}
	\caption{The overall system}
	\label{fig:overallsystem}
\end{figure}

Figure \ref{fig:overallsystem} presents a logical architecture of the mixed reality system. It is composed by three main subsystems: (i) the gaming subsystem responsible of managing the game application; (ii) the motion capture subsystem responsible of tracking patient's movement; and (iii) the display subsystem responsible of displaying the virtual game environment in a physical environment.

\subsection{The gaming subsystem}

\begin{figure}[h]
	\centering
\includegraphics[scale=0.7]{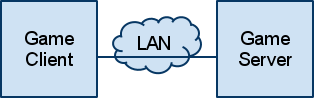}
	\caption{The game subsystem}
	\label{fig:gamesubsystem}
\end{figure}
As shown by figure \ref{fig:gamesubsystem} the gaming subsystem follows a client-server pattern:
\begin{itemize}
\item The game server: The mechanics of the game is implemented at this level. This means that the client sends user's inputs to the server following a specific protocol; the game server translates these inputs to game events that are used to compute the next state of the game according to the game mechanics specifications. Once the new state of the game has been computed, update events are then send to game client(s) to update their local copies of game objects which causes the update their graphical aspect.
\item The game client: This component fulfills the following functions: (i) receiving patient's movement events from the motion capture system; (ii) forwarding these events to the game server; (iii) receiving events from the game server to update game objects states; and finally (iv) generating game's outputs in terms of 3D scenes and sounds.
\end{itemize}

\subsection{Motion capture subsystem}

\begin{figure}[h]
	\centering
		\includegraphics[scale=0.7]{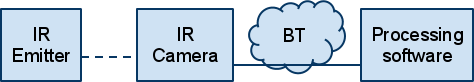}
	\caption{The motion capture schema}
	\label{fig:motioncapture}
\end{figure}

The motion capture subsystem is responsible of tracking patient's hand and to communicate these movements to the game client. As shown by Figure \ref{fig:motioncapture} the motion capture system is composed of three components:
\begin{itemize}
\item IR emitter: this is an active sensor built using a simple infrared (IR) emitter that is attached on the patient's hand using a Velcro strap. This basic device is very cheap (less than 1 euro) and convenient to use even with patients that suffer from spasticity.
\item IR camera: To track the IR emitter a Nintendo Wiimote has been used as an IR camera. The Nintendo Wiimote is an affordable device (less than 30 euros) that has been used in many open source projects. Consequently, a plethora of open source libraries are now available to use Wiimote as pointing device.
\item A processing software: the role of this component is to translate Wiimote events into mouse events so that the couple composed of IR emitter and IR camera will be considered from the operating system point of view as a standard pointing device.
\end{itemize}

\subsection{Game display subsystem}
Within a mixed reality context, it is necessary to display game objects directly on the physical world. Thanks to a new generation of mini video projectors, namely pico projectors, it is possible to display the game on a planar surface almost anywhere. Within the context of this experiment, the pico projector is attached to a support to display the game on a standard table.
The pico projectors are affordable (less than 200 euros), lightweight (less that 300 gr); and easy to used simply by plugging a VGA cable on a computer. However, they are limited in terms of brightness (less than 30 lumens for the one that has been used in this experiment). This constraint has to be taken into account when building the game graphical environment to select colors and contrasts to compensate this limitation.
\par\medskip
\textbf{The overall system:}
\begin{figure}[h]
	\centering
\begin{tabular}{cc}
		\includegraphics[scale=0.7]{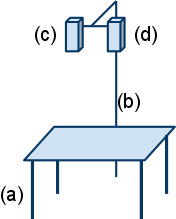}&\includegraphics[scale=0.4]{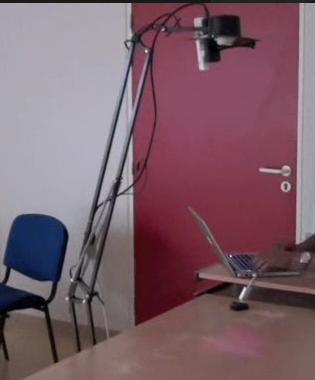}
\end{tabular}
	\caption{The overall system \citep{fishtank}}
	\label{fig:overallsystemschema}
\end{figure}

Finally the overall system is presented in Figure \ref{fig:overallsystemschema}.  The patient sits in front of the table (a) with the IR emitter device attached to his hand. The IR camera (c) and pico projector (d) are placed on the top of a support (b). A laptop computer is used to run the game client and to establish a Bluetooth connection with the WiiMote. A video demonstrating the system can be seen on \citep{fishtank}.

\subsection{Virtual Scenario}
\label{game}

\begin{figure}[h]
	\centering
		\includegraphics[scale=0.7]{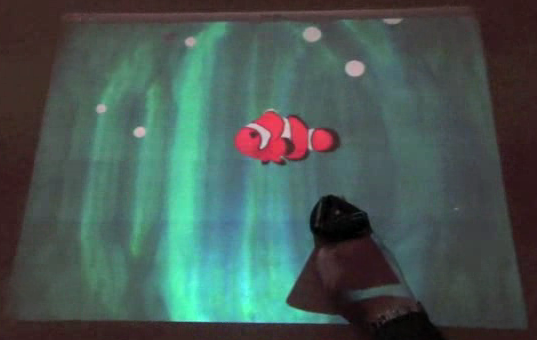}
	\caption{The thank fish game}
	\label{fig:fishgame}
\end{figure}

The virtual scenario that has been used in this study consists of a fish character that navigates in a fish tank (Figure \ref{fig:fishgame}). The goal of the player is to reach and touch the character. To avoid non-voluntary collisions with the fish, whenever the patient touches the fish a progression bar is displayed. The touching event is endorsed only when the progression bar has been completely filled.
The fish character has the following behaviors:

\begin{itemize}
\item wander: when this behavior is started the fish is asked to move around without any specific goal. This causes the character to explore randomly a 3D sphere with predefined radius.
\item pursue: this behavior causes the character to pursue a specific 3D object. For instance, the patient's hand is represented as an object within the 3D world and the clown fish can pursue the patient's hand when asked to.
\item flee: this behavior causes the character to flee a specific 3D object. As in the pursue case, the character can flee patient's hand when asked to.
\end{itemize}

The virtual character is controlled by a software agent representing its AI (Artificial Intelligence). The software agent is executed by the game server. It is given as inputs patient's hand position and velocity; according to its strategy, the software agent asks the clown fish character to adopt one of the three behaviors, namely: wander, pursue or flee. In this study a basic game difficulty adjustment strategy has been implemented.
In order to manage patient's frustration, a dynamic game difficulty adjustment process is required. This process is indented to modify game parameters and game's entities behavior to adapt the difficulty of the game to patient by observing its actions and deducing his/her capabilities.
Dynamic game balancing is an important feature addressed by our work on therapeutic games for post stroke rehabilitation. However, since this is not the primary subject of this study only a brief description of the difficulty adjustment strategy is presented.

\begin{figure}[h]
	\centering
		\includegraphics[scale=0.7]{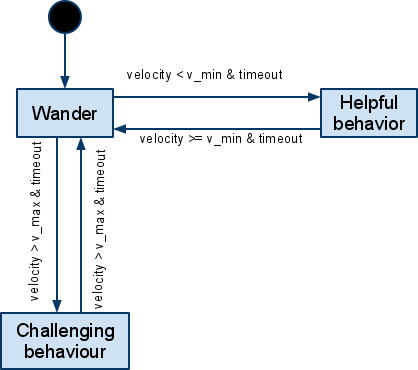}
	\caption{State if the agent implementing strategy}
	\label{fig:statemachine}
\end{figure}

The software agent controlling the fish game character is responsible of implementing the game difficulty adjustment strategy.
The internal state of the software agent is a finite state machine presented in Figure \ref{fig:statemachine}. The states of the FSM that can be described as follow:
\begin{itemize}
\item wander: this is the initial state where the agent asks the fish character to simply move randomly
\item helpful behavior: when the software agent observes, for a certain period, that the patient hand's velocity is under $v_{min}$ threshold -- this is interpreted as the patient is stuck -- then the fish is asked to adopt a helpful behavior by pursuing patient's hand.
\item challenging behavior: when the patient hand's velocity exceeds, for a certain time, a defined threshold ($v_{max}$) -- this is interpreted as the patient is doing too well -- then the software agent asks the fish to challenge the patient by fleeing from his/her hand.
\end{itemize}
\section{Protocol}
\label{protocol}
\subsection{Participants}
The mixed reality system described above has been tested on 3 therapists (two male and one female, ages 20-30). As described in section \ref{methods}, during  this experiment therapists  took at the same time the roles of patient and  therapist.
\par \medskip

\subsection{The Experiment}
The experiment was composed of different sessions, each testing a different 'form of therapy' (classical therapy, computer games, mixed reality games). Each session took 15 minutes. After the last session a double questionnaire was conducted adding another 15 minutes. \par \medskip
Summarizing, the session were:
\begin{enumerate}
\item Classical rehabilitation (CR): a simple classical rehabilitation session, using a exercise which is not related to any form of gaming
\item	 Playing a game on the computer (PC)
\item	 Playing a game in mixed reality (MR)
\end{enumerate}

The game on the computer and in the mixed reality were identical. Order of sessions was randomized. 
\par \medskip

Since the aim was to compare alternative tools for rehabilitation, it was necessary to define a typical set of tasks that could be represented by all tools. Participants were then asked to replicate the following exercise with all the tools:\\
From the bottom right : Try to reach the object\\
From the bottom left  : Try to reach the object\\
From the upper right  : Try to reach the object\\
From the upper left : Try to reach the object\\
For the PC game and the MR system the object to reach was a the fish character. 

\par \medskip

\subsubsection{Patient's Questionnaire}
At the end of the three sessions, experiences were reported in two ways. Firstly  therapists were asked to  report about their experience when acting like a patient. In this case they had to   answer to the In-game version of the GEQ \citep{GEQ07} -  which is a concise version of the core questionnaire - basing on  what they think a patient would answer. The Game Experience Questionnaire (GEQ)  assesses game experience as scores on seven components: Immersion, Flow, Competence, Positive and Negative Affect, Tension, and Challenge.  In-game GEQ has an identical component structure and consists of items selected from the main module but is composed of only 14 questions. Only two items are used for every component. Items for each component are listed below. 

\begin{itemize}
\item \emph{Competence} (feeling successful and/or skilled)
\item \emph{Sensory and Imaginative Immersion} (feeling interested and/or impressed)
\item  \emph{Flow} (forgetting everything around you and/or feeling completely absorbed)
\item \emph{Tension} (feeling frustrated and/or feeling irritable)
\item \emph{Challenge} (feeling challenged and/or put a lot of effort)
\item \emph{Negative affect} (feeling bored and/or feeling tired)
\item \emph{Positive affect} (feeling contend and/or feeling good)
\end{itemize}

\par\medskip

Before entering in detail into the results some clarification about the above used terms had to been done.
In psychology, \emph{affect} is an emotion or subjectively experienced feeling. 
Positive affects are feelings such as enjoyment, interest, excitement. Negative affects are feelings such as anger, rage, disgust, fear.
 \emph{Flow} \citep{csik} is the mental state of operation in which a person in an activity is fully immersed in a feeling of energized focus, full involvement, and success in the process of the activity.

\par \medskip

\subsubsection{Therapist's Questionnaire}
Second part of the assessment was addressed to therapists as therapists. They were asked to answer questions related to how funny, useful and usable the different systems were. Finally, using open-ended questions, participants were also asked
to justify their choice and to comment on main advantages and problems for each system. 

\subsubsection{Data analysis method}
\label{dataanalysis}
Since this study was conducted as a pilot study only descriptive statics are used. Results are reported as (Mean $\pm$ Standard Deviation) All statistical analysis was performed using R (http://www.r-project.org) version 2.12.0

\section{Results}
\label{results}
\subsection{Therapists as patients questionnaire}
\label{resultspatient}
\begin{figure}[h]
	\centering
		\includegraphics[scale=0.7]{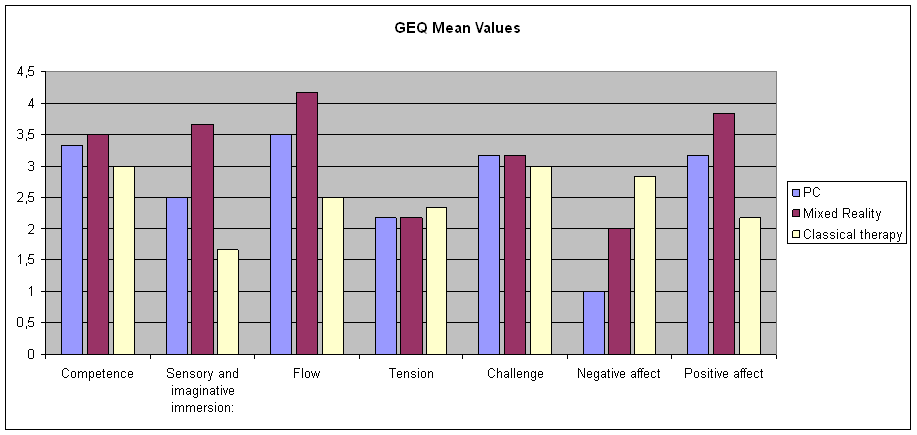}
	\caption{Graphical representation of the mean data}
	\label{fig:graph}
\end{figure}

First of all we will note the results of the In-game version of the GEQ on all the three 'therapists as patients'. In Table \ref{fig:fuma} mean and standard deviation for the questionnaire data are represented. Figure 	\ref{fig:graph} gives a graphical representation of mean data.


\begin{table}[h]
\caption{Mean and standard deviation for the questionnaire data}
\center
\label{fig:fuma}
\begin{tabular}{|l|c|c|c|}
\hline
Component & PC & Mixed Reality & Classical Therapy \\
\hline
Competence & 	($3.34 \pm 0.74$) & 	($3.5 \pm 0.5$) & 	($3 \pm 1.15$) \\
Immersion & 	($2.5 \pm 1.11$) & 	($3.67 \pm 0.94$) & 	($1.67 \pm 0.74$) \\
Flow & 		($3.5 \pm 1.11$) & 	($4.17 \pm 1.06$) & 	($2.5 \pm 0.95$)\\
Tension & 		($2.17 \pm 1.21$) & 	($2.17 \pm 1.06$) & 	($2.33 \pm 1.10$) \\
Challenge & 	($2.17 \pm 0.68$) & 	($3.17 \pm 1.06$) & 	($3 \pm 0.81$) \\
Negative affect&	($1 \pm 0$) & 		($2 \pm 1$) & 		($2.84 \pm 1.21$) \\
Positive affect &	($3.17 \pm 0.68$) & 	($3.84 \pm 0.37$) & 	($2.17 \pm 0.68$) \\
\hline
\end{tabular}
\end{table}

Always in their patient's role therapist were asked to answer the following question: "In general, which therapy did you preferred to do? (Classify exercises in order of preference)".	
This means that we asked them to order the therapies as they thought the patients will do.
All the three gave the following order: 1)Mixed reality game,  2)Pc game,	3)Physical therapy.

\subsection{Therapists as therapists questionnaire}
\label{resultstherapists}
Hereafter the results of the therapists' questionnaire (without any order and with duplicates). In synthesis they were asked as therapists to list the most negative and positive aspects of all the used systems.
\par\medskip
\emph{List of the most negative aspect(s)}:\\				
For the pc game: \\
- The patient had to look at the screen while playing	
\\- The  patient had to grab the mouse			
\\ - The  fact to use only the upper part of the triangle was upsetting (here the therapist is referring to the game)			
\\ - The  screen was small			
\\- It's difficult for the patient to move the mouse 			
\\ - It's difficult for the patient to use the mouse			
\\
For the MR game: 	 \\ - it takes more time to start up  (here the therapist is referring to the system/wii calibration)	
\\
For classical therapy:
\\ - Tiresome and painful			
\\ - Motivation is low			
\\ - The patient has to confront with his own limitation			
				
\par\medskip
\emph{List the most positive aspect(s)}: 	\\	
For the pc game:\\ - Games are fun
	\\ - The patient can forget his environment
	\\ - Different and may be funny
\par\medskip
For the MR game:  \\ - Accessible. You can use whatever surface you want (for example a table)
	\\ - Not so difficult to use as in the mouse case
	\\ - Less material needed once started 
\\
None of the therapists gave particular comments on the classical therapy.
\par\medskip
Finally, as  therapists they had to answer the following question: "In general, which therapy did you preferred to do? (Classify exercises in order of preference)".			
This time they were thus asked their own preferences on  the therapies.
All the three gave the following order: 1)Mixed reality game,  2)Pc game,	3)Physical therapy.

\section{Discussion}
\label{discussion}
As described in Section \ref{methods} the main goal of the experiment is to compare a mixed reality system (MRS) with two alternative single-user tools through an pilot study. The two alternative tools were classical physical rehabilitation and an ad-hoc post stroke pc game. We expected the MRS and the PC game to be more accepted  -- in Shackel's sense \citep{Shackel1991} -- than classical rehabilitation when therapists are playing the patient role. The MRS easiness of use is also expected to be higher than the PC one. We are then analyzing perceived utility, usability, likeability (affective evaluation) (see Figure \ref{fig:shackel})

It is important to note that this is a pilot study involving a limited number of participants ($n=3$) to prepare a larger scale experiment. Consequently, quantitative data are presented as a support for discussion and no statistically valid generalizations could be inferred at this stage.

\subsection{Evaluating possible patient's involvement}
\label{patient}
As we can see from Figure \ref{fig:graph} elements linked to the environment oblivion (immersion and flow) have different ratings. 

Flow is rated low for classical rehabilitation ($2.5 \pm 0.95$) while PC ($3.5 \pm 1.11$) and mixed reality ($4.17 \pm 1.06$) have higher scores. 
Sensory and imaginative immersion ratings are definitely lower for classical rehabilitation ($1.67 \pm 0.74$), next comes PC ($2.5 \pm 1.11$) and finally mixed reality ($3.67 \pm 0.94$)

We can conclude that within the context of this experiment, both MRS and PC have been considered as more immersive, followed by classical therapy. 

This is an expected result since it is already known that using games for rehabilitation creates a sense of meaningfulness for repeated movements and encourages flow experience by forgetting everything around and being focused only on the task to be performed \citep{Reid2004}.

Lack of tension is an element characterizing tiredness and boringness. Results show that this component is almost similar for PC, MRS and classical therapy with respectively ($2.17 \pm 1.21$), ($2.17 \pm 1.06$) and ($2.33 \pm 1.10$). 

So, within the context of this experiment all systems have been considered equivalent in terms of tiredness and boringness. When asked about this point, therapists highlight the fact that post stroke rehabilitation demands a lot of effort from the patient. So, game based rehabilitation can facilitate, to some extent, the context of rehabilitation but patients will sooner or later show fatigue and pain and the therapeutic session has to stop.

If we look at negative affects, PC ($1 \pm 0$) is the one rating lower, followed by mixed reality ($2 \pm 1$) and the classical therapy ($2.84 \pm 1.21$) is the one rated with higher negative affect. 

When asked qualitatively about this fact, therapists mentioned the problem of exertion when using MRS. In fact, there is a one to one mapping of patient movements between the physical and virtual world. So the scaling and artificial compensation of movements is not possible as it could be performed within a virtual environment by changing speed and sensitivity of the pointing device for instance. This was considered as a very interesting remark that we were not aware of before conducting the study. Consequently, within MRS the exertion of the patient is an important challenge to take into account and innovative solutions have to be provided to compensate patient's limited movement such as modifying dynamically size of objects or moving objects closer to patient when she is stuck.

What is very interesting is that while MR is seen as a possible cause of moderately negative affects, it is rated as the systems with the higher positive affects ( ($3.84 \pm 0.37$) for MR; ($3.17 \pm 0.68$) for PC;  ($2.17 \pm 0.68$) for classical therapy)

This reinforces the idea that MR has the potential to increase the volume of rehabilitation by increasing the exercises they do. However, as said previously, this should be done while taking in account patient's exertion.

Finally, competences and challenges are rated in a very similar way. This means that the effort the 'patient' had to put to understand and perform exercises was the same. 

To summarize, when playing the role of patients, therapists have considered:
 \begin{itemize}
\item that PC, MRS and classical therapy were equivalent on competence, tension and challenge components.
\item Both PC and MRS systems are better than classical therapy on immersion, flow, negative affect and positive affects.
\item MRS has been considered as worse than PC on negative affect due to worries expressed on fatigue. 
\end{itemize}

\subsection{Evaluating therapist's acceptance}
\label{Therapist}
If we switch to the therapists questionnaire we can describe why the MRS has a higher potential to be accepted by patients. 

In fact, one of the most underlined problems with PC rehabilitation was the use of the mouse as a pointing device. This limitation becomes one of potentiality of MRS (see the comment: Not difficult to use as in the mouse case). In addition, comments on PC usage are also related to limitations about screen dimension. Finally the most interesting comment is linked to the fact that the patients have to look in another place (the PC screen) in order to perform the task. This could be a great problem when using this kind of device on some patients. The MRS on the other hand can project on any surface. 

Finally it is interesting to note that all three subjects gave the same order of preferences: (1) MRS, (2) PC, (3) Classical therapy both as patients and as therapists.

\section{Conclusion and future works}
\label{conclusion}

Objective of this paper was to present a mixed reality system (MRS) for rehabilitation of the upper limb after stroke. The system answers the following challenges:
(i) increase motivation of patients by making the training a personalized experience 
(ii) take into account patients impairments by offering intuitive and easy to use interaction modalities 
(iii) make it possible to therapists to track patient's activities and to evaluate/track patient's progress  (iv) open opportunities to telemedicine and tele rehabilitation 
(v) provide an economically acceptable system by reducing both equipment and management costs. 

It's our opinion that it is important to evaluate the acceptance of these systems not only from the the patient's point of view but also from therapists' point of view. We decide then to test this system wih therapists in a pilot study conducted in conjunction with a French hospital. The main assumption behind this experiment was that, even if clinical efficiency is demonstrated, if therapist does not accept the system it will be used only for academic studies and may not be widely accepted in real rehabilitation centers.  The pilot described in this paper involved 3 therapists who 'played the role' of  patients. The main idea was that, because of the years they spent with patients, they are able to 'simulate' patients' reactions to the system.
Three sessions, one using conventional rehabilitation, another using an ad hoc developed game on a PC, and another using a mixed reality version of the same game were held.  We expected the MRS and the PC game to be accepted more than physical rehabilitation by the therapists in the patient's role(H1), and MRS easiness of use to be considered higher than the PC one(H2). 

Results shows that while the effort the 'patient' has to put in the exercises was practically the same for all the systems,  Mixed Reality can be a mean to help patients to forget exertion and potentially augment the number of exercises they do.
We can say that Mixed Reality and PC games requires the same amount of effort than classical rehabilitation. However, because of the characteristics of Mixed Reality and PC games we can suppose that patients will prefer to use this two systems with a preference for the MRS (H2).
In addition, analyzing the therapists questionnaire we can say that the Mixed Reality System has a higher potential to be accepted by patients (H1). 
In fact, one of the most underlined problems with PC rehabilitation was the use of a mouse as a device. This limitation becomes one of potentiality of Mixed reality. In addition the MRS can be project in whatever surface and avoid the problem to look in a place and perform the task in another. 

The pilot study  described in this paper was not intended to deliver statistical evidence, but simply to give approximate values guiding the set-up of following experiments. 
For this reason we scheduled two more experiments. 
The first one is an in depth version of the pilot study presented in this paper. For this experiment we want to gather a greater number of data using the same 'role playing' protocol. 
The second one will 'drop' the role playing and will analyze actual stroke rehabilitation of the upper limb with patients and therapists each in they own role. 
The main idea behind this two experiment is to compare  results from the two in order to understand the discrepancy between patients attitudes and therapists expectations towards the system.

\section{Acknowledgements}
\label{acknowledgements}
 Will be added

\end{document}